\documentstyle[12pt]{article}
\renewcommand{\baselinestretch}{1}
\begin{document}
\title{Nonleptonic charmed meson decays:\\ Quark diagrams and final-state
interactions}
\author{P. \.{Z}enczykowski\\
\\
Department of Theoretical Physics\\
The H. Niewodnicza\'nski Institute of Nuclear Physics\\
Radzikowskiego 152, Krak\'ow, Poland\\
}
\maketitle
\begin{abstract}
Effects of final-state interactions in nonleptonic decays of charmed
mesons are studied in the framework of
quark-diagram approach.
For the case of $u$-$d$-$s$ flavour symmetry
we discuss how the inelastic
coupled-channel rescattering effects
(and, in particular, resonance formation in the final state) modify
the input quark-diagram weak amplitudes.
It is shown that such inelastic effects
lead to the appearance of nonzero relative phases between various quark
diagrams, thus invalidating some of the conclusions drawn in the past within
the diagrammatical approach. The case of SU(3) symmetry-breaking in
Cabibbo once-forbidden $D^0$ decays is also studied
in some detail.
\end{abstract}

\newpage
\baselineskip = 10 pt
\renewcommand{\baselinestretch}{1.685}
\small
\normalsize

\section{Introduction}
Various theoretical models of
nonleptonic decays of charmed mesons have been developed over the years.
The most general and complete one is the diagrammatical approach of
Chau and Cheng
\cite{Chau86,Chau87,Chau89}.
The factorization method \cite{BWS} is just a special case of
this approach.  Another subset of diagrammatical approach is singled out
by large-$N_c$ arguments \cite{Lee}.  The basic problem with the
diagrammatical approach is the way in which final-state interactions
(FSI) and SU(3) breaking are treated.
The importance of FSI has been stressed by
Lipkin \cite{Lipkin},
Sorensen \cite{SorD23},
Kamal and Cooper \cite{Kamal},
Donoghue \cite{Don86},
Chau \cite{Chau83},
Chau and Cheng \cite{ChCh92},
Hinchliffe and Kaeding \cite{Hinchcli},
and others.

Complete descriptions of nonleptonic decays {\em must} take into
account
final-state interactions. Since full dynamical calculations of these
effects are
not possible at present, a meaningful comparison of theoretical models with
experiment requires at least a phenomenological estimate of FSI.
Such an approximate estimate may be obtained using eg. unitarity
constraints \cite{Sorensen}.  Alternatively, one may consider
approaches
based on approximate flavour-symmetry groups. Their predictions
include automatically
{\em all} effects of those FSI which are invariant under these symmetries
\cite {Lipkin}.
The diagrammatical methods of Chau and Cheng provide an approach
complementary to that based on flavour-symmetry group \cite{Lipkin,Hinchcli}.
It has been argued \cite{Don86} that "it is folly to proceed with the
quark-diagram approach without considering these (i.e. rescattering)
effects" since "rescattering can mix up the classification of
diagrams".
Although the latter statement is obviously true, one should
realize that the quark-diagram approach - being complementary to
that based on flavour symmetry groups -
"deals
with effective quark diagrams with all FSI included" \cite{Chau87}.

\newpage
The aim of this paper is

1) to examine in some detail
in what way the introduction of flavour-symmetric FSI in the form of
coupled-channel
rescattering effects renormalizes the input quark-diagram weak
amplitudes (with particular emphasis on
resonance formation in the final state)
and to compare the results thus obtained with the treatment of FSI
adopted in the diagrammatical approach so far
(Section 3),
and

2) to discuss some aspects of
flavour-symmetry breaking (especially $D^0$ decays into $\pi \pi$ and
$K \overline{K}$, (Section 4)).

\section{General}
 In this paper we will consider how weak decay
amplitudes are changed when final-state rescattering effects are added.
When there is only one possible final state, the answer is well-known
and given by Watson's theorem \cite{Watson}.
 For the coupled-channel case the situation was discussed in
\cite{Babelon} where the generalized Watson's theorem was introduced.
For our purposes we will use the $K$-matrix parametrization of the
time-reversal-invariant $S$-matrix \cite{Don86}:
\begin{equation}
\label{eq:Kmatrix}
S=\frac{1+iK}{1-iK}
\end{equation}
with $K$ real and symmetric.
Let the charmed meson $D$ decay weakly into $n$ two-body (in general
many-body) coupled channels $j$.  The $S$-matrix may then be written
as
\begin{equation}
\label{eq:weakS}
S = \left[
\begin{array}{cc}
1+O(W^2) & iW^T \\
iW       & S_0
\end{array}
\right]
\end{equation}
where $S_0$ is an $n \times n$ submatrix describing purely strong
interaction in the coupled channel system, while $iW$ is an
$n$-dimensional vector describing weak $D$-decays into these two-body
channels.
Since $S$ is unitary, $W$ contains all final-state strong effects
as well.
The corresponding reaction matrix $K$ may be written as
\begin{equation}
\label{eq:weakK}
K =
\left[
\begin{array}{cc}
0 & w^T \\
w & K_0
\end{array}
\right]
\end{equation}
where $K_0$ is the $K$-matrix for the $n \times n$ purely strong
submatrix $S_0 = (1+iK_0)/ (1-iK_0)$, while $w$ is an $n$-dimensional
vector describing FSI-unmodified weak decays of $D$ mesons.

The relation between the input and the full weak decay amplitudes
is then
\begin{equation}
\label{eq:genWatson}
W = (1+S_0)w = \frac{2}{1-iK_0}w
\end{equation}
i.e. in the case of vanishing FSI we have
\begin{equation}
\label{eq:genWatson0}
W = 2w
\end{equation}
Matrix equation (\ref{eq:genWatson}) admixes into a given decay amplitude
contributions from all coupled channels.  For the $n=1$ case
one has $S_0 = \exp (2i\delta _0)$, ($K_0 = \tan \delta_0$) and
Eq.(\ref{eq:genWatson}) reproduces Watson's theorem:
\begin{equation}
\label{eq:Watson}
W = 2 w \cos \delta _0 \exp (i\delta _0)
\end{equation}
For the coupled-channel case Eq.(\ref{eq:genWatson}) constitutes a
generalized case of Watson's theorem.
It includes all final-state interactions: both elastic
scattering
and {\em all} inelastic rescattering effects.
As we shall see in this paper explicitly,
such inelastic rescattering effects have not been included
in the diagrammatical approach of refs. \cite{Chau86,Chau87,Chau89} so far.
For reasons related to the introduction of FSI the diagrammatical approach has
been criticised
by Donoghue \cite{Don86}, and, very recently, by Hinchcliffe and Kaeding
\cite{Hinchcli}.
The question of how the inelastic effects change the overall picture
of quark-line approach was first examined in ref.\cite{Don86}.
Here we will show how the problem gets simplified conceptually, and -
in the most important case of final-state $q \overline{q}$ resonance
contribution
- also computationally,
through diagonalization of the $K_0$ ($S_0$)
matrix and application of Watson's theorem in the diagonalizing
basis.
Diagonalization will be discussed on the example of
Cabibbo-allowed parity-conserving decays of $D^0$ and $D^+_s$
into the $PV$ (pseudoscalar meson + vector meson) final states.
Results of a similar treatment of the
Cabibbo-allowed parity-violating decays $D^0$, $D^+_s \rightarrow PP$
will also be presented.
Finally, we will consider the interesting case of Cabibbo-forbidden
parity-violating decays of $D^0$ into  the $\pi \pi$ and $K \overline{K}$
channels where SU(3) breaking is expected to play a significant role.
In all of our examples we will neglect complications due to possible
presence of coupled channels other than those listed above.

\section{Cabibbo-allowed decays of $D^0$ and $D^+_s$ into PV and PP
final states}
\subsection{Parity conserving PV decays}
{\bf 3.1.1. $D^0 \rightarrow PV$.}
The parity-conserving part of weak interactions induces $D^0$ decays
into eight possible final p-wave $PV$ channels:
$ K^- \rho ^+$, $\overline{K}^0 \rho ^0$, $\overline{K}^0 \omega$,
$\eta _s
\overline{K}^{*0}$
and
$K^{*-} \pi^+$, $\overline{K}^{*0} \pi ^0$, $\overline{K}^{*0} \eta_{ns}$,
$\phi \overline{K}^{0} $.

In the first four decay channels (below called $\underline{P}V$)
the strange quark from the weak decay of the charmed quark
ends up in the pseudoscalar meson $P$,
while in the latter four decay channels (called $\underline{V}P$) this
strange quark ends up in the vector meson.
The decays proceed through diagrams $(a)$, $(b)$, $(c)$ from
Fig. 1.

\newpage
\setlength {\unitlength}{1.6pt}

\begin{picture}(260,340)

\put(20,240){\begin{picture}(80,90)
\put(0,55){\vector(1,0){15}}
\put(15,55){\line(1,0){15}}
\put(0,45){\line(1,0){15}}
\put(30,45){\vector(-1,0){15}}
\put(30,55){\vector(2,1){20}}
\put(50,65){\line(2,1){20}}
\put(30,45){\line(2,1){20}}
\put(70,65){\vector(-2,-1){20}}
\multiput(30,55)(2,-2){8}{\circle*{1.2}}
\put(50,40){\oval(10,10)[l]}
\put(50,45){\vector(2,-1){10}}
\put(60,40){\line(2,-1){10}}
\put(50,35){\line(2,-1){10}}
\put(70,25){\vector(-2,1){10}}
\put(35,10){$(a)$}
\put(0,60){$c$}
\end{picture}}

\put(20,140){\begin{picture}(80,90)
\put(0,55){\vector(1,0){15}}
\put(15,55){\line(1,0){15}}
\put(0,45){\line(1,0){15}}
\put(30,45){\vector(-1,0){15}}
\multiput(30,55)(0,-2){6}{\circle*{1.2}}
\put(30,55){\vector(2,1){20}}
\put(50,65){\line(2,1){20}}

\put(30,45){\line(2,-1){20}}
\put(70,25){\vector(-2,1){20}}

\put(40,50){\line(2,1){15}}
\put(70,65){\vector(-2,-1){15}}
\put(40,50){\vector(2,-1){15}}
\put(55,42.5){\line(2,-1){15}}
\put(35,10){$(c)$}
\put(0,60){$c$}
\end{picture}}

\put(120,240){\begin{picture}(80,90)
\put(0,55){\vector(1,0){25}}
\put(25,55){\line(1,0){15}}
\put(0,45){\line(1,0){25}}
\put(40,45){\vector(-1,0){15}}
\multiput(41,55)(2,-1){5}{\circle*{1.2}}
\put(40,55){\vector(2,1){20}}
\put(60,65){\line(2,1){20}}

\put(40,45){\line(2,-1){20}}
\put(80,25){\vector(-2,1){20}}

\put(50,50){\line(2,1){15}}
\put(80,65){\vector(-2,-1){15}}
\put(50,50){\vector(2,-1){15}}
\put(65,42.5){\line(2,-1){15}}
\put(35,10){$(b)$}
\put(0,60){$c$}
\end{picture}}

\put(20,40){\begin{picture}(80,90)
\put(0,55){\vector(1,0){10}}
\multiput(11,55)(2,0){5}{\circle*{1.2}}
\put(15,55){\oval(10,10)[b]}
\put(20,55){\line(1,0){10}}
\put(0,45){\line(1,0){15}}
\put(30,45){\vector(-1,0){15}}
\put(30,55){\vector(2,1){20}}
\put(50,65){\line(2,1){20}}

\put(30,45){\line(2,-1){20}}
\put(70,25){\vector(-2,1){20}}

\put(40,50){\line(2,1){15}}
\put(70,65){\vector(-2,-1){15}}
\put(40,50){\vector(2,-1){15}}
\put(55,42.5){\line(2,-1){15}}
\put(35,10){$(e)$}
\put(0,60){$c$}
\end{picture}}

\put(120,140){\begin{picture}(90,90)
\put(0,55){\vector(1,0){15}}
\put(15,55){\line(1,0){10}}
\put(0,45){\line(1,0){15}}
\put(25,45){\vector(-1,0){10}}
\put(25,50){\oval(10,10)[r]}
\put(40,50){\vector(2,1){25}}
\put(40,50){\line(2,-1){25}}
\put(65,62.5){\line(2,1){25}}
\put(90,25){\vector(-2,1){25}}
\put(90,65){\vector(-2,-1){15}}
\put(90,35){\line(-2,1){15}}
\put(75,57.5){\line(-2,-1){15}}
\put(60,50){\vector(2,-1){15}}
\multiput(31,50)(2,0){5}{\circle*{1.2}}
\put(35,10){$(d)$}
\put(0,60){$c$}
\end{picture}}

\put(120,40){\begin{picture}(90,90)
\put(0,55){\vector(1,0){15}}
\put(15,55){\line(1,0){10}}
\put(0,45){\line(1,0){15}}
\put(25,45){\vector(-1,0){10}}
\put(25,50){\oval(10,10)[r]}
\put(40,50){\vector(2,1){25}}
\put(40,50){\line(2,-1){25}}
\put(65,62.5){\line(2,1){25}}
\put(90,25){\vector(-2,1){25}}
\put(90,65){\vector(-2,-1){15}}
\put(90,35){\line(-2,1){15}}
\put(75,57.5){\line(-2,-1){15}}
\put(60,50){\vector(2,-1){15}}
\multiput(20,46)(0,2){5}{\circle*{1.2}}
\put(35,10){$(f)$}
\put(0,60){$c$}
\end{picture}}

\put(5,20){Fig. 1~ Quark-line diagrams for weak meson decays }
\put(5,10){\phantom{Fig. 1~} $(a)$, $(b)$ - factorization; $(c)$ -
$W$-exchange;
$(d)$ - annihilation;}
\put(5,0){\phantom{Fig. 1~} $(e)$ - "horizonthal penguin"; $(f)$ - "vertical
penguin"}
\end{picture}

Let us first consider the case when there are no final-state strong
interactions ($S_0 =1$).
By $a,b,c$ ($a',b',c'$) we denote matrix elements corresponding to
diagrams
$(a),(b),(c)$, for $\underline{P}V$ ($\underline{V}P$) channels
respectively.
Evaluation of contributions from these diagrams yields the following
weak decay amplitudes
\begin{eqnarray}
\label{eq:D0a}
2<(\overline{K}\rho )_{3/2}|w|D^0> &=& -\frac{1}{\sqrt{3}}(a+b)
\nonumber \\
2<(\overline{K}\rho )_{1/2}|w|D^0> &=& \frac{1}{\sqrt{6}}(b-2a-3c)
\nonumber \\
2<\overline{K}^0 \omega |w|D^0>      &=& -\frac{1}{\sqrt{2}}(b+c)
\nonumber \\
2<\eta _s \overline{K}^{*0}|w|D^0>    &=& -c
\end{eqnarray}
and
\begin{eqnarray}
\label{eq:D0b}
2<(\overline{K}^* \pi )_{3/2}|w|D^0> &=& -\frac{1}{\sqrt{3}}(a'+b')
\nonumber \\
2<(\overline{K}^* \pi )_{1/2}|w|D^0> &=& \frac{1}{\sqrt{6}}(b'-2a'-3c')
\nonumber \\
2<\overline{K}^{*0} \eta _{ns}|w|D^0>                  &=&
-\frac{1}{\sqrt{2}}(b'+c') \nonumber
\\
2<\phi \overline{K}^0|w|D^0>    &=& -c'
\end{eqnarray}
Subscripts $1/2$ and $3/2$ denote total isospin of $\overline{K} \rho$
and $\overline{K}^* \pi$ states.
Dependence on Cabibbo factors is suppressed both in Eqs (\ref{eq:D0a},
\ref{eq:D0b}) and elsewhere in this paper.
That is, all expressions on the right-hand side of Eqs (\ref{eq:D0a},
\ref{eq:D0b}) are to be multiplied by the Cabibbo factor of $\cos ^2
\Theta _C$ : $ -(a+b)/\sqrt{3} \rightarrow - \cos ^2 \Theta _C
\cdot (a+b)/\sqrt{3}$ etc.
The normalisation is such that matrix elements $a,b,c$ etc. are
identical with those used in \cite{Chau86,Chau87,Chau89}
(weak decay amplitudes in the case of vanishing FSI
are given by Eq.(\ref{eq:genWatson0})).

Final-state interactions may include quark-exchange
\cite{Don86}, resonance formation \cite{BWS}, elastic scattering etc.
Of these it is resonance formation that is in general expected to affect
naive approaches most significantly \cite{Lipkin,SorD23,Kamal,Don86}.
In the following
we will consider FSI through $q\overline{q}$ resonance formation only.
Later, it should become clear that our {\em general} conclusions remain valid
also when other FSI are included.
The final-state coupled-channel processes to be considered are
visualised in Fig.2


\begin{picture}(170,80)
\put(70,10){\begin{picture}(70,70)
\put(0,70){\line(2,-1){30}}
\put(30,55){\vector(1,0){5}}
\put(35,55){\line(1,0){5}}
\put(40,55){\line(2,1){30}}

\put(0,60){\line(2,-1){20}}
\put(20,50){\line(0,-1){5}}
\put(20,40){\vector(0,1){5}}
\put(20,40){\line(-2,-1){20}}

\put(0,20){\line(2,1){30}}
\put(30,35){\line(1,0){5}}
\put(40,35){\vector(-1,0){5}}
\put(40,35){\line(2,-1){30}}

\put(70,30){\line(-2,1){20}}
\put(50,40){\line(0,1){5}}
\put(50,50){\vector(0,-1){5}}
\put(50,50){\line(2,1){20}}
\end{picture}}
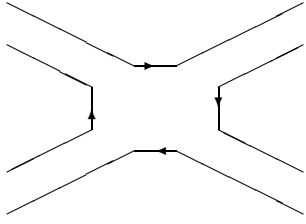
\put(20,0){Fig.2 Resonance contribution to final-state interactions}

\end{picture}

Since the intermediate state is a pseudoscalar state (with kaon flavour
quantum numbers), the flavour structure of the $K_0$ matrix may
be easily calculated from the product of two $F$-type ($VP\rightarrow
P'$) couplings.
(In general, one chooses
symmetric(antisymmetric) D(F) coupling
$Tr(M_1\{M_2,M_3\})$ ($Tr(M_1[M_2,M_3])$)
when the product
of charge conjugation parities of the three mesons is positive (negative).)

This gives the following strong $K_0$ matrix for the
$\underline{P}V$ subsector:
\begin{equation}
\label{eq:K0}
K_0 =
\left[
\begin{array}{ccc}
\frac{3}{2} & \frac{\sqrt{3}}{2} & \sqrt{\frac{3}{2}} \\
\frac{\sqrt{3}}{2} & \frac{1}{2} & \frac{1}{\sqrt{2}} \\
\sqrt{\frac{3}{2}} & \frac{1}{\sqrt{2}} & 1
\end{array}
\right]
\cdot \kappa ^2
\end{equation}
where the channels are ordered
$(\overline{K}\rho )_{1/2}$, $\overline{K}^0 \omega $, $\eta_s
\overline{K}^{*0}$
and "$\kappa$" is the flavour-symmetric strength factor of the
$VP \rightarrow P'$ coupling.
(In the $(\overline{K}\rho)_{3/2}$ sector the $K_0$ matrix
obviously vanishes).

The eigenvalues and eigenvectors of matrix (\ref{eq:K0}) are:
\begin{eqnarray}
\label{eq:eigen}
\lambda _1 = & 3 \kappa ^2 \equiv \tan \delta _1~~~~~ & |1> =
\frac{1}{\sqrt{6}}(\sqrt{3}(\overline{K}\rho)_{1/2}+(\overline{K}^0
\omega)+\sqrt{2}(\eta_s \overline{K}^{*0})) \nonumber
\\
\lambda _2 = & 0 \equiv \tan \delta _2~~~~~           & |2> =
\frac{1}{\sqrt{6}}(-\sqrt{3}(\overline{K}\rho)_{1/2}
+(\overline{K}^0 \omega)+\sqrt{2}(\eta_s \overline{K}^{*0}))
\nonumber
\\
\lambda _3 = & 0 \equiv \tan \delta _3~~~~~           & |3> =
\frac{1}{\sqrt{3}}(\sqrt{2}(\overline{K}^0 \omega)-(\eta_s
\overline{K}^{*0}))
\end{eqnarray}

Equations (\ref{eq:D0a}) may be easily rewritten in the new basis
(for the $I=1/2$ sector):
\begin{eqnarray}
\label{eq:D0anew}
2<1|w|D^0> & = & -\frac{1}{\sqrt{3}}(a+3c) \nonumber \\
2<2|w|D^0> & = & \frac{1}{\sqrt{3}}(a-b) \nonumber \\
2<3|w|D^0> & = & -\frac{1}{\sqrt{3}}b
\end{eqnarray}
In this basis Eq.(\ref{eq:genWatson}) reads
\begin{equation}
\label{eq:Watsondiagonal}
<j|W|D^0> = 2 \cos \delta _j \exp (i \delta _j) <j|w|D^0>
\end{equation}
with $j= 1,2,3$. Eq.(\ref{eq:Watsondiagonal}) is as simple as the standard
one-channel case of Watson's theorem (Eq.(\ref{eq:Watson})).
Going back from Eqs.\ref{eq:Watsondiagonal} to the old basis (and
adding the expression for the $I=3/2$ decay amplitude) one obtains
\begin{eqnarray}
\label{eq:D0A}
<(\overline{K}\rho )_{3/2}|W|D^0> & = & -\frac{1}{\sqrt{3}}(A+B)
\nonumber \\
<(\overline{K}\rho )_{1/2}|W|D^0> & = & \frac{1}{\sqrt{6}}(B-2A-3C)
\nonumber \\
<\overline{K}\omega|W|D^0> & = & -\frac{1}{\sqrt{2}}(B+C) \nonumber \\
<\eta _s \overline{K}^{*0}|W|D^0> & = & -C
\end{eqnarray}
with
\begin{eqnarray}
\label{eq:FSI1}
A & = & a \nonumber \\
B & = & b \nonumber \\
C & = & c+\left( c+\frac{a}{3}\right) (\cos \delta _1 \exp
i\delta _1 - 1)
\end{eqnarray}
where vanishing of $\delta _2$ and $\delta _3$ has been taken into account.

From Eqs (\ref{eq:D0A},\ref{eq:FSI1}) we see that the case with
resonance-induced coupled-channel effects differs from
the no-FSI case by a change in the size and phase of the $C$ parameter
{\em only}:
\begin{eqnarray}
\label{eq:change}
C = c & \rightarrow & C =  c+ \left( c+\frac{a}{3}
\right)(\cos \delta _1 \exp i\delta _1 - 1)
\end{eqnarray}
Thus, after including resonance-induced coupled-channel effects, the
reduced
matrix elements corresponding to diagrams $(a)$ and $(b)$ remain unchanged
and real, while the matrix element of diagram $(c)$ acquires new size
and {\em nonvanishing} (relative to $(a)$ and $(b)$) {\em phase}. The
coupled-channel
effects may generate a sizable nonvanishing effective $(c)$-type diagram
even if the original $(c)$-type amplitude was negligible as assumed in many
papers \cite{Don86,BWS,SorD23}. This is shown
in diagrammatic terms in Fig.3. Eq.(\ref{eq:change}) is a mathematical
representation of how FSI-modified quark-diagram amplitudes are obtained.
When rescattering reactions proceed by quark exchange, the corresponding
$K_0$ matrix has three nonzero eigenvalues. The counterparts of Eqs
(\ref{eq:FSI1}) are then less transparent and more complicated leading
in particular to the appearance of Zweig-rule violating
"hairpin" diagrams \cite{Chau89a}. In this paper we will
not consider these quark-exchange contributions in detail.

In ref. \cite{Don86} a simple case of flavour symmetry
breaking was considered. While the couplings were still assumed to be
$SU(3)$-symmetric, symmetry breaking was introduced through
flavour-symmetry-breaking phase-space factor.  Such breaking modifies
the reaction matrix $K_0$ in the same way, independently of whether the
rescattering is
due to quark exchange or resonance formation. In our case Eq.(\ref{eq:K0})
is replaced by
\begin{equation}
\label{eq:K0epsilon}
K_0 =
\left[
\begin{array}{ccc}
\frac{3}{2} & \frac{\sqrt{3}}{2} & \sqrt{\frac{3}{2}}\epsilon \\
\frac{\sqrt{3}}{2} & \frac{1}{2} & \frac{1}{\sqrt{2}}\epsilon \\
\sqrt{\frac{3}{2}}\epsilon & \frac{1}{\sqrt{2}}\epsilon & \epsilon ^2
\end{array}
\right]
\cdot \kappa ^2
\end{equation}
where $\epsilon $ is a measure of the phase-space-induced suppression of
the contribution from the $\eta _s \overline{K}^{*0}$ channel.
By writing Eq.(\ref{eq:K0epsilon}) with a universal $\kappa $ we assume
an artificial situation in which the (pseudoscalar)
resonances $P'$ themselves are still SU(3)-symmetric.
Although this assumption is certainly not expected in an approach in
which pseudoscalars from the intermediate $PV$ states exhibit $SU(3)$
symmetry breaking, it is not incorrect on general grounds and
- for illustrative purposes -
may be used as a simplifying assumption.

The eigenvalues of the $K_0$ matrix are now $\lambda _{1,2,3} =
(2+\epsilon ^2) \kappa ^2, 0, 0$. Proceeding as before we obtain
formulae
(\ref{eq:D0A}) again,
where the only change with respect to Eqs (\ref{eq:FSI1},\ref{eq:change}) is
\begin{eqnarray}
\label{eq:changeepsilon}
C = c & \rightarrow & C =  c+ \left( c+\frac{a}{2+\epsilon ^2}
\right)(\cos \delta _1 \exp i\delta _1 - 1)
\end{eqnarray}
As before, only the size and phase of the $(c)$-type amplitude is affected
which should be obvious from Fig.3.
\\

\begin{picture}(260,120)

\put(0,40){\begin{picture}(70,75)
\put(0,55){\vector(1,0){15}}
\put(15,55){\line(1,0){15}}
\put(0,45){\line(1,0){15}}
\put(30,45){\vector(-1,0){15}}
\put(30,55){\vector(2,1){20}}
\put(50,65){\line(2,1){20}}
\put(30,45){\line(2,1){20}}
\put(70,65){\vector(-2,-1){20}}
\multiput(30,55)(2,-2){8}{\circle*{1.2}}
\put(50,40){\oval(10,10)[l]}
\put(50,45){\vector(2,-1){10}}
\put(60,40){\line(2,-1){10}}
\put(50,35){\line(2,-1){10}}
\put(70,25){\vector(-2,1){10}}
\put(35,10){$(a)$}
\put(0,60){$c$}
\end{picture}}

\multiput(72.5,65)(0,4){13}{\line(0,1){2}}

\put(75,40){
\begin{picture}(80,75)
\put(0,75){\line(2,-1){30}}
\put(30,60){\vector(1,0){5}}
\put(35,60){\line(1,0){5}}
\put(40,60){\line(2,1){30}}

\put(0,65){\line(2,-1){20}}
\put(20,55){\line(0,-1){5}}
\put(20,45){\vector(0,1){5}}
\put(20,45){\line(-2,-1){20}}

\put(0,25){\line(2,1){30}}
\put(30,40){\line(1,0){5}}
\put(40,40){\vector(-1,0){5}}
\put(40,40){\line(2,-1){30}}

\put(70,35){\line(-2,1){20}}
\put(50,45){\line(0,1){5}}
\put(50,55){\vector(0,-1){5}}
\put(50,55){\line(2,1){20}}

\end{picture}
}

\put(145,90){\vector(1,0){15}}

\put(165,40){\begin{picture}(80,90)
\put(0,55){\vector(1,0){15}}
\put(15,55){\line(1,0){15}}
\put(0,45){\line(1,0){15}}
\put(30,45){\vector(-1,0){15}}
\multiput(30,55)(0,-2){6}{\circle*{1.2}}
\put(30,55){\vector(2,1){20}}
\put(50,65){\line(2,1){20}}

\put(30,45){\line(2,-1){20}}
\put(70,25){\vector(-2,1){20}}

\put(40,50){\line(2,1){15}}
\put(70,65){\vector(-2,-1){15}}
\put(40,50){\vector(2,-1){15}}
\put(55,42.5){\line(2,-1){15}}
\put(35,10){$(c)$}
\put(0,60){$c$}
\end{picture}}

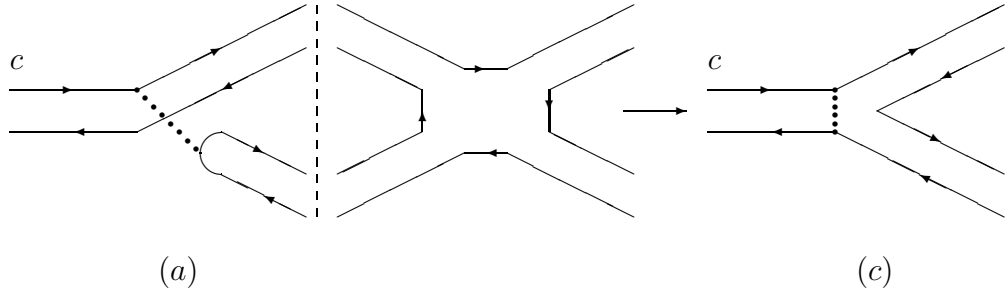
\put(0,15){Fig. 3~ Resonance-induced generation of $(c)$-type (exchange)
amplitude}
\put(0,5){\phantom{Fig. 3~} from
$(a)$-type (factorization) diagram}

\end{picture}

The procedure
applied above to the $\underline{P}V$ sector may be repeated in the
$\underline{V}P$ sector.
In this sector the $K_0$
matrix has the form given in Eq.(\ref{eq:K0})
(with decay channels ordered $(\overline{K}^*\pi )_{1/2}$,
$\overline{K}^{*0}\eta _{ns}$, $\phi \overline{K}^0$) and the same is true
for the off-diagonal ($\underline{P}V-\underline{V}P$) part of the total
$K_0$ matrix.
Introducing three eigenvectors $|j'>$ ($j'=1,2,3$) of the $\underline{V}P$
sector (given by formulae (\ref{eq:eigen}) with
$(\overline{K}\rho )_{1/2}$, $\overline{K}^0 \omega $, $\eta _s
\overline{K}^{*0}$ replaced by
$(\overline{K}^*\pi )_{1/2}$, $\overline{K}^{*0}\eta _{ns}$, $\phi
\overline{K}^0$
respectively
one obtains the total $K_0$ matrix:
\begin{equation}
K_0=\left(
\begin{array}{cccc}
3 & 3 & 0 & ... \\
3 & 3 & 0 & ... \\
0 & 0 & 0 & ... \\
. & . & . & ... \\
. & . & . & ...
\end{array}
\right) \cdot \kappa ^2
\end{equation}
with rows and columns ordered $|1>,|1'>$, $|2>,|2'>$ ... .
Diagonalizing the total $K_0$ matrix and repeating the procedure
outlined earlier one obtains - in addition to Eqs (\ref{eq:D0A}) - the
following expressions for the new FSI-modified
$\underline{V}P$ decays:
\begin{eqnarray}
\label{eq:D0B}
<(\overline{K}^*\pi )_{3/2}|W|D^0> & = & -\frac{1}{\sqrt{3}}(A'+B')
\nonumber \\
<(\overline{K}^*\pi )_{1/2}|W|D^0> & = & \frac{1}{\sqrt{6}}(B'-2A'-3C')
\nonumber \\
<\overline{K}^{*0}\eta_{ns}|W|D^0> & = & -\frac{1}{\sqrt{2}}(B'+C')
\nonumber \\
<\phi \overline{K}^0|W|D^0> & = & -C'
\end{eqnarray}
The reduced matrix elements of Eqs (\ref{eq:D0A},\ref{eq:D0B}) are now given
by
\begin{eqnarray}
\label{eq:changefull}
A    & = & a  \nonumber \\
A'   & = & a' \nonumber \\
B    & = & b  \nonumber \\
B'   & = & b' \nonumber \\
C-C' & = & c-c' \nonumber \\
C+C' & = &
c+c'+\left(\cos \delta ^0_{PV} \exp i\delta ^0_{PV} - 1 \right)
(c+c'+\frac{a+a'}{3})
\end{eqnarray}
where $\tan \delta ^0_{PV} = 6 \kappa ^2$.
In conclusion, including resonance-induced coupled-channel effects in the
SU(3)-symmetric case results in a change of size and phase of matrix
elements of diagrams $(c)$ {\em only}. Since in the diagrammatical
approach the size of
$(c)$-type amplitudes was treated as a free parameter
\cite{Chau86,Chau87,Chau89} anyway, the only observable
effect of FSI is the appearance of (in general different) nonzero
phases of
$C$ and $C'$.

In refs.\cite{Chau86,Chau87,Chau89} an attempt to
include the effects of FSI has been made. To permit easy comparison with
expressions derived above we rewrite below a few formulae
from Table 1 of ref.\cite{Chau89} ($SU(3)$-symmetry case):
\begin{eqnarray}
\label{eq:Chau89}
<(\overline{K}\rho )_{3/2}|W|D^0> &=& \frac{1}{\sqrt{3}} ({\cal A}+{\cal
B}) \exp i\delta ^{K \rho }_{3/2}\nonumber \\
<(\overline{K}\rho )_{1/2}|W|D^0> &=& \frac{1}{\sqrt{6}} (2{\cal A}-{\cal
B}+3{\cal C}) \exp i\delta ^{K\rho }_{1/2}\nonumber \\
<(\overline{K}^*\pi )_{3/2}|W|D^0> &=& \frac{1}{\sqrt{3}} ({\cal A'}+{\cal
B'}) \exp i\delta ^{K^*\pi }_{3/2}\nonumber \\
<(\overline{K}^*\pi )_{1/2}|W|D^0> &=& \frac{1}{\sqrt{6}} (2{\cal A'}-{\cal
B'}+3{\cal C'}) \exp i\delta ^{K^*\pi }_{1/2}\nonumber \\
<\phi \overline{K}^0|W|D^0>        &=& {\cal C'} \exp i\delta ^{\phi K}
\nonumber \\
<\overline{K}^{*0}\eta _{ns}|W|D^0>&=& (mixture)
\end{eqnarray}
In papers \cite{Chau86,Chau87,Chau89}
amplitudes ${\cal A}$, ${\cal B}$, ${\cal C}$, etc.
(Eq.(\ref{eq:Chau89})) were real while
the phase factors were allowed both  real and
imaginary parts in the hope of taking
into account all inelasticities and phase shifts possible.
After comparing Eq.(\ref{eq:Chau89}) with
Eqs(\ref{eq:D0A},\ref{eq:D0B},\ref{eq:changefull})
we see that resonance contribution does {\em not} follow
the ansatz of \cite{Chau87,Chau89}: In Eq.(\ref{eq:Chau89}) the
{\em relative} phases of contributions from diagrams $(a)$,$(b)$,$(c)$
are {\em zero}, while proper consideration of resonance-induced
inelastic
coupled-channel effects leaves relative phases of $A,B,A',B'$ zero, but
adds an important {\em nonzero} phase to $C$ and $C'$. Thus,
analysis of ref.\cite{Chau87,Chau89} does {\em not} take into
account effects due to the possible formation of resonances in the
final state {\em even in the case of SU(3)-symmetry}
(though elastic scattering is taken care of). (See also
the paper of Hinchcliffe and Kaeding \cite{Hinchcli} for a general
comment on the inclusion of FSI in the diagrammatic approach.)

\newpage
{\bf 3.1.2. $D^+_s \rightarrow PV$.}
  In this case one obtains the following FSI-modified expressions:
\begin{eqnarray}
\label{eq:Ds}
<(\rho \pi )_2|W|D^+_s> & = & 0    \nonumber \\
<(\rho \pi )_1|W|D^+_s> & = & D-D' \nonumber \\
<\omega \pi ^+|W|D^+_s> & = & -\frac{1}{\sqrt{2}}(D+D') \nonumber \\
<\phi \pi ^+|W|D^+_s> & = & -A' \nonumber \\
<\rho ^+ \eta _{ns}|W|D^+_s> & = & -\frac{1}{\sqrt{2}}(D+D')
\nonumber \\
<\rho ^+ \eta _s|W|D^+_s> & = & -A \nonumber \\
<K^* \overline{K}^0|W|D^+_s> & = & -B-D' \nonumber \\
<\overline{K}^{*0} K^+|W|D^+_s> & = & -B'-D
\end{eqnarray}
with
\begin{eqnarray}
\label{eq:changeDs}
A   & = & a  \nonumber \\
A'  & = & a' \nonumber \\
B   & = & b  \nonumber \\
B'  & = & b' \nonumber \\
D+D'& = & d+d' \nonumber \\
D-D'& = & d-d' +(\cos \delta ^{+s}_{PV} \exp i \delta ^{+s}_{PV} - 1)
(d-d' - \frac{1}{3}(b-b'))
\end{eqnarray}
From Eqs(\ref{eq:Ds},\ref{eq:changeDs}) we see that, as before, matrix
elements corresponding
to diagrams $(a)$, $(b)$ are not affected by FSI, while for
$(d)$-type diagrams it is only the difference $D-D'$ that is modified.
Even if one starts with $d-d'=0$ (expected on the basis of flavour
$u \leftrightarrow \overline{d}$ symmetry, see Fig. 1), the
coupled channel effects generate a nonvanishing effective $D-D'$
proportional to $b-b'$.

\subsection{Parity-violating PP decays}
In the parity-violating $s$-wave decays $ D^0,D^+_s \rightarrow PP $
the final-state interactions are most probably dominated by formation of
scalar resonances \cite{Don80}.
We accept here that in the $S=-1$ sector under discussion
the properties of these resonances are susceptible to a simple
$q\overline{q}$ description (see next section for discussion of this
assumption) so that the treatment of the previous section may be applied.
Proceeding as before one can
then derive:
\begin{eqnarray}
\label{eq:D0}
<(\overline{K}\pi )_{3/2}|W|D^0> & = & -\frac{1}{\sqrt{3}}(A+B)
\nonumber \\
<(\overline{K}\pi )_{1/2}|W|D^0> & = & \frac{1}{\sqrt{6}}(B-2A-3C)
\nonumber \\
<\overline{K}^0\eta _{ns}|W|D^0>   & = & -\frac{1}{\sqrt{2}}(B+C)
\nonumber \\
<\eta _s \overline{K}^0|W|D^0> & = & -C
\end{eqnarray}
with $A=a$, $B=b$ and $C=c+(c+a/3)(\cos \delta ^0_{PP} \exp i \delta
^0_{PP}-1)$.
Even if the input quark-model $(c)$-type amplitude is negligible
\cite{SorD23}, the effective $W$-exchange amplitude may be significant.

Similarly, for $D^+_s$ decays we get
\begin{eqnarray}
<\pi ^+ \pi ^0|W|D^+_s> & = & 0 \nonumber \\
<\pi ^+ \eta _{ns}|W|D^+_s> & = & -\sqrt{2} D \nonumber \\
<\pi ^+ \eta _s|W|D^+_s> & = & - A \nonumber \\
<K^+ \overline{K}^0|W|D^+_s> & = & -(B+D)
\end{eqnarray}
with $A=a$, $B=b$ and $D=d+(\cos \delta ^{+s}_{PP} \exp \delta ^{+s}_{PP} -
1)(d+b/3)$.

Examples of Cabibbo-allowed $D^0$
and $D^+_s$ decays studied in this and previous subsections show how
quark-line-diagram approach is affected when
$q\overline{q}$ resonance formation in the final state is taken into
account. In phenomenological approaches in which sizes
of matrix elements constitute free parameters, the only observable
effect of such resonances in the final state is the appearance
of {\em non-zero phases of the $(c)$ and $(d)$} - type amplitudes (when
compared with the FSI-unaffected amplitudes $(a)$ and $(b)$).
This is a part of the
quark-diagram version of the general statement that "predictions based
on flavour-symmetry groups automatically include {\em all} effects of
those FSI which are invariant under these symmetries" \cite{Lipkin}.
It should be obvious now that in the most general case the
amplitudes corresponding to individual diagrams should be allowed {\em
independent nonzero phases}.
Depending on what types of FSI are considered one can then have various
conditions imposed upon these phases.

\section{Cabibbo-forbidden $D^0$ decays and $SU(3)$ symmetry breaking}
      In this section
we will analyze in some detail the case of $SU(3)$ symmetry breaking
in Cabibbo-once-forbidden $D^0$ decays (in particular, the long-standing
problem of the
$\Gamma (D^0 \rightarrow K^+ K^- )$/ $\Gamma ( D^0 \rightarrow \pi ^+ \pi
^-)$ ratio).
In these decays there are 9 possible $s$-wave $PP$ final states:
$\pi ^+ \pi ^-$, $\pi ^0 \pi ^0$, $K^- K^+$, $K^0 \overline{K}^0$,
$\eta _8 \pi ^0$, $\eta _1 \pi ^0$, $\eta _8 \eta _8$,
$\eta _8 \eta _1$, $\eta _1 \eta _1$.
Each of these can be reached from the initial $D^0$ state by appropriate
linear combination of six reduced matrix elements corresponding to
diagrams $(a)$-$(f)$ of Fig. 1.
Omitting the Cabibbo factor of $\sin \Theta _C \cos \Theta _C$
(see comment after Eqs (\ref{eq:D0a},\ref{eq:D0b}))
we obtain the following
expressions for FSI-uncorrected weak decays of $D^0$:
\begin{eqnarray}
\label{eq:D0forbidden}
2<\pi ^+ \pi^-|w|D^0>&=&-(a+c-e-2f) \nonumber \\
2<\pi ^0 \pi ^0|w|D^0>&=&-\frac{1}{\sqrt{2}}(b-c+e+2f)\nonumber \\
2<K^- K^+|w|D^0>&=&-(\tilde{a}+\tilde{c}+e+2f) \nonumber \\
2<K^0 \overline{K}^0|w|D^0>&=&(-c+\tilde{c}+2f) \nonumber \\
2<\pi ^0 \eta _8|w|D^0>&=&\frac{1}{\sqrt{3}}(\tilde{b}-c-e) \nonumber \\
2<\pi ^0 \eta _1|w|D^0>&=&-\frac{1}{\sqrt{6}}(\tilde{b}+2c+2e) \nonumber \\
2<\eta _8 \eta _8|w|D^0>&=&\frac{1}{\sqrt{2}}(b-c-\frac{1}{3}e-2f)
+\frac{2}{3\sqrt{2}}(2(c-\tilde{c})+(\tilde{b}-b)) \nonumber \\
2<\eta _8 \eta _1|w|D^0>&=&\frac{1}{\sqrt{2}}(b+2c-\frac{2}{3}e)
+\frac{1}{3\sqrt{2}}(-4(c-\tilde{c})+(\tilde{b}-b)) \nonumber \\
2<\eta _1 \eta _1|w|D^0>&=&-\sqrt{2}(\frac{1}{3}e+f)
+\frac{\sqrt{2}}{3}(c-\tilde{c}+b-\tilde{b})
\end{eqnarray}
In Eq.(\ref{eq:D0forbidden}) matrix elements with (without) a tilde
correspond to a strange (nonstrange) pair emitted
in the original weak interaction process.

In standard approaches the contribution from diagrams $(e)$ and $(f)$
is negligible \cite{ChCh92} (it vanishes in the $SU(3)$ limit).
Accordingly, we will neglect $(e)$, $(f)$-type amplitudes
in FSI-uncorrected $D^0$ decays.
In the $SU(3)$ limit we also have $\tilde{a}=a$, $\tilde{b}=b$, $\tilde{c}=c$.
Our aim is to study if and how
final-state interactions
reintroduce $(e)$- and $(f)$- type amplitudes and lift
equalities $\tilde{a}=a$, $\tilde{b}=b$, $\tilde{c}=c$.

We will consider the contribution of $PP$ coupled-channel effects only.
The two-meson $s$-wave $PP$ state may interact strongly through
formation of neutral scalar resonances $S$. Assuming these belong to
a $q\overline{q}$ nonet (see discussion later on) we consider three
resonances with flavour quantum numbers of
$\pi ^0$, $\eta _8$, and $\eta _1$.
Their couplings to the $PP$-state are of $D$-type.

The $K_0$-matrix splits block-diagonally into three submatrices in the
isospin $I=2,1,0$ sectors respectively:
\begin{enumerate}
\item \underline{sector} $\underline{I=2}$: , $K_0(I=2)=0$
\\
for the $(\pi
\pi)_{I=2}$ state only.
\item \underline{sector} $\underline{I=1}$

\begin{equation}
\label{eq:KI1}
K_0(I=1) = 2
\left[
\begin{array}{ccc}
1 & \sqrt{\frac{2}{3}} & \frac{2}{\sqrt{3}} \\
\sqrt{\frac{2}{3}} & \frac{2}{3} & \frac{2\sqrt{2}}{3} \\
\frac{2}{\sqrt{3}} & \frac{2\sqrt{2}}{3} & \frac{4}{3}
\end{array}
\right]
\cdot (\kappa _S^{I=1})^2
\end{equation}
with rows (columns) corresponding to states
$(K\overline{K})_{I=1}$, $\pi ^0 \eta _8$, $\pi ^0 \eta _1$.

The eigenvalues and their corresponding eigenvectors are:
\begin{eqnarray}
\label{eq:eigenforb1}
\lambda _1^{I=1} =& 6 \kappa ^2_S& \equiv \tan \delta ^{I=1}_1
\nonumber \\
&|1^{I=1}>& = \frac{1}{3}(\sqrt{3}(K\overline{K})_{I=1} + \sqrt{2} (\pi ^0
\eta _8) + 2 (\pi ^0 \eta _1))\nonumber \\
\lambda _2^{I=1} =&0& \equiv \tan \delta ^{I=1}_2
\nonumber \\
&|2^{I=1}>& =\frac{1}{3}(\sqrt{6}(K\overline{K})_{I=1}-(\pi ^0 \eta _8)
-\sqrt{2} (\pi ^0 \eta _1))\nonumber \\
\lambda_ 3^{I=1} =&0 \equiv & \tan \delta ^{I=1}_3
\nonumber \\
&|3^{I=1}>& =-\frac{1}{\sqrt{3}} (\sqrt{2}(\pi ^0 \eta _8) - (\pi ^0 \eta _1))
\end{eqnarray}
Vanishing of $\lambda _2^{I=1}$ results from the assumed
nonet symmetry of couplings, which relates the couplings of
$(K\overline{K})_{I=1}$, $(\pi ^0 \eta _8)$, and $(\pi ^0 \eta _1)$
in such a way that linear combination $|2^{I=1}>$ decouples from
the scalar $I=1$, $I_z=0$ meson (hereafter denoted
$\pi
^0_S$).
State $|2^{I=1}>$ would couple to $\pi ^0_S$ if $SU(3)$-breaking
coupling $Tr(M_{\pi ^0_S}\{M_2 \lambda _8 M_3+M_3 \lambda _8 M_2\})$
were introduced.
Vanishing of $\lambda _3^{I=1}$ corresponds to the absence of hairpin
diagrams ($|3^{I=1}> = \pi ^0 \eta _s$).

\item \underline{sector} $\underline{I=0}$
\begin{equation}
\label{eq:KI0}
K_0(I=0) = 2
\left[
\begin{array}{ccccc}
3&-\sqrt{3}&1&2&0 \\
-\sqrt{3}&3&-\sqrt{3}&0&0 \\
1&-\sqrt{3}&1&0&0 \\
2&0&0&2&0 \\
0&0&0&0&0
\end{array}
\right] (\kappa _S^{I=0})^2
\end{equation}

with rows (columns) corresponding to states (in that order):
$(K\overline{K})_{I=0}$, $(\pi \pi)_{I=0}$,
$\eta _{ns} \eta _{ns}$, $\eta _s \eta _s$, $\eta _{ns} \eta _s$.

The eigenvalues and eigenvectors are:
\begin{eqnarray}
\label{eq:eigenforb0}
\lambda _1^{I=0} =& 12 \kappa ^2_S \equiv & \tan \delta _1^{I=0}
\nonumber \\
&|1^{I=0}>&=\frac{1}{3}(-2(K\overline{K})_{I=0}-(\eta _s \eta _s)+\sqrt{3}
(\pi \pi)_{I=0}-(\eta _{ns} \eta _{ns}))\nonumber \\
\lambda _2^{I=0} =& 6 \kappa ^2_S \equiv & \tan \delta _2^{I=0}
\nonumber \\
&|2^{I=0}>&=\frac{1}{3}((K\overline{K})_{I=0}+2(\eta _s \eta _s)+\sqrt{3}
(\pi \pi)_{I=0}-(\eta _{ns} \eta _{ns}))\nonumber \\
\lambda _3^{I=0} =& 0 \equiv & \tan \delta _3^{I=0}
\nonumber \\
&|3^{I=0}>&=\frac{1}{3}(2(K\overline{K})_{I=0}-2(\eta _s \eta _s)+
\frac{\sqrt{3}}{2}
(\pi \pi)_{I=0}-\frac{1}{2}(\eta _{ns} \eta _{ns}))\nonumber \\
\lambda _4^{I=0} =& 0 \equiv &\tan \delta _4^{I=0}
\nonumber \\
&|4^{I=0}>&=\frac{1}{2}(
(\pi \pi)_{I=0}+\sqrt{3}(\eta _{ns} \eta _{ns}))\nonumber \\
\lambda _5^{I=0} =& 0 \equiv & \tan \delta _5^{I=0}
\nonumber \\
&|5^{I=0}>&= \eta _{ns} \eta _{s}
\end{eqnarray}
State $|1^{I=0}>$ is $SU(3)$ singlet, while $|2^{I=0}>$ is $SU(3)$ octet.
The couplings of the remaining three states are zero as a result
of the quark-level nonet symmetry of D-type coupling. For example,
$\lambda _4^{I=0}$ may be nonzero if
$(\pi \pi)_{I=0}$ and $(\eta _{ns} \eta _{ns})$ couplings to
$\sigma _{ns}$ (i.e. to $(u\overline{u}+d\overline{d})/\sqrt{2}$)
are {\em not} related as specified by D-type
coupling. Similarly, deviation of the scale of couplings involving strange
quarks
from those in which strange quarks are absent would result in nonvanishing of
the coupling between the $|3^{I=0}>$ state and the $I=0$ mesons.
Coupling of state $|5^{I=0}>$ to $q\overline{q}$ mesons would be
nonzero if hairpin diagrams were allowed.
\end{enumerate}

Let us reexpress the general FSI-modified formulae of type
(\ref{eq:D0forbidden}) in the $K_0$-matrix approach:
\\

a) \underline{sector} $\underline{I=2}$
\\
\begin{eqnarray}
\label{eq:FSIforb2}
<(\pi \pi )_{I=2}|W|D^0>&=&-\frac{1}{\sqrt{3}}(A+B) = \nonumber \\
 &=& 2<(\pi \pi)_{I=2}|w|D^0> = (-\frac{1}{\sqrt{3}})(a+b)
\end{eqnarray}
b) \underline{sector} $\underline{I=1}$
\\
\begin{eqnarray}
\label{eq:FSIforb1}
<(K\overline{K})_{I=1}|W|D^0>&=&-\frac{1}{\sqrt{2}}(\tilde{A}+C+E) =
\nonumber \\
& =& \frac{2}{\sqrt{3}}\cdot T^{(1)}_1 <1^{I=1}|w|D^0>
+ 2<(K\overline{K})_{I=1}|w|D^0> \nonumber \\
<\pi ^0 \eta _8|W|D^0>&=&\frac{1}{\sqrt{3}}(\tilde{B}-C-E) =
\nonumber \\
&=& \frac{2\sqrt{2}}{3}\cdot T^{(1)}_1 <1^{I=1}|w|D^0>
+ 2 <\pi ^0 \eta _8|w|D^0>   \nonumber \\
<\pi ^0 \eta _1|W|D^0>&=&-\frac{1}{\sqrt{6}}(\tilde{B}+2C+2E) =
\nonumber \\
&=& \frac{4}{3}\cdot T^{(1)}_1 <1^{I=1}|w|D^0>
+ 2 <\pi ^0 \eta _1|w|D^0>
\end{eqnarray}
where
\begin{equation}
\label{eq:1SU3}
2<1^{I=1}|w|D^0> =
-\frac{1}{\sqrt{6}}(\tilde{a}+3c+3e)\stackrel{SU(3)}{\rightarrow}
-\frac{1}{\sqrt{6}}(a+3c)
\end{equation}
\begin{equation}
\label{eq:KK1}
2<(K\overline{K})_{I=1}|w|D^0>= -\frac{1}{\sqrt{2}}(\tilde{a}+c+e)
\stackrel{SU(3)}{\rightarrow} -\frac{1}{\sqrt{2}}(a+c)
\end{equation}
and
\begin{equation}
\label{eq:Tdef}
T^{(1)}_1 \equiv \cos \delta ^{I=1}_1 \exp i\delta ^{I=1}_1 -1
\end{equation}
Solution of Eqs (\ref{eq:FSIforb1}) for SU(3)-symmetric input amplitudes is
\begin{eqnarray}
\label{eq:FSIsol1}
  \tilde{A} & = & a \nonumber \\
  \tilde{B} & = & b \nonumber \\
  C+E &  =  & c+T^{(1)}_1 (c+\frac{a}{3})
\end{eqnarray}
\\
c) \underline{sector} $\underline{I=0}$
\begin{eqnarray}
\label{eq:FSIforb0}
<(K\overline{K})_{I=0}|W|D^0>&=&
\frac{1}{\sqrt{2}}(-\tilde{A}+C-2\tilde{C}-E-4F)=
\nonumber \\
&=&\frac{2}{3}(-2T^{(0)}_1<1^{I=0}|w|D^0>+T^{(0)}_2<2^{I=0}|w|D^0>)
\nonumber \\
&&+2<(K\overline{K})_{I=0}|w|D^0>
\nonumber \\
<(\pi \pi)_{I=0}|W|D^0>&=&
\sqrt{\frac{2}{3}}(-A+\frac{1}{2}B-\frac{3}{2}C+\frac{3}{2}E+3F)=
\nonumber \\
&=&\frac{2}{\sqrt{3}}(T^{(0)}_1<1^{I=0}|w|D^0>+T^{(0)}_2<2^{I=0}|w|D^0>)
\nonumber \\
&&+2<(\pi \pi)_{I=0}|w|D^0>
\nonumber \\
<\eta _8 \eta _8|W|D^0>&=&
\frac{1}{\sqrt{2}}(B-C-\frac{1}{3}E-2F)+\frac{2}{3\sqrt{2}}
(2(C-\tilde{C})+(\tilde{B}-B))=
\nonumber \\
&=&\frac{2}{3}(-T^{(0)}_1<1^{I=0}|w|D^0>+T^{(0)}_2<2^{I=0}|w|D^0>)
\nonumber \\
&&+2<\eta _8 \eta _8|w|D^0>
\nonumber \\
<\eta _1 \eta _1|W|D^0>&=&
-\frac{\sqrt{2}}{3}(-C+\tilde{C}-B+\tilde{B}+E+3F)=
\nonumber \\
&=&-\frac{2}{3}T^{(0)}_1<1^{I=0}|w|D^0>+2<\eta _1 \eta _1|w|D^0>
\nonumber \\
<\eta _1 \eta _8|W|D^0>&=&
\frac{1}{\sqrt{2}}(B+2C-\frac{2}{3}E+
\frac{4}{3}(\tilde{C}-C)+\frac{1}{3}(\tilde{B}-B))= \nonumber \\
&=&-\frac{4}{3}T^{(0)}_2<2^{I=0}|w|D^0>+2<\eta _1 \eta _8|w|D^0>
\end{eqnarray}
where
\begin{eqnarray}
\label{eq:neww}
2<1^{I=0}|w|D^0> &=&-\frac{\sqrt{2}}{3}(a-\tilde{a}+3(c-\tilde{c})-3e-9f)
\nonumber \\
2<2^{I=0}|w|D^0>&=&-\frac{1}{\sqrt{2}}(\frac{1}{3}
(\tilde{a}+2a)+c+2\tilde{c}-e)
\end{eqnarray}
and
\begin{equation}
\label{eq:T2}
T^{(0)}_j = \cos \delta ^{I=0}_j \exp \delta ^{I=0}_j - 1
\end{equation}
with $j=1,...5$ (in Eq.(\ref{eq:FSIforb0}) we have used $T^{(0)}_{3,4,5}=0$).

Eqs (\ref{eq:FSIforb0}) are much simplified when one accepts the
$SU(3)$-limit for input weak decays:
\begin{eqnarray}
\label{eq:SU3lim}
2<1^{I=0}|w|D^0>
& \stackrel{SU(3)}{\rightarrow}& 0 \nonumber \\
2<2^{I=0}|w|D^0>
&\stackrel{SU(3)}{\rightarrow}&-\frac{1}{\sqrt{2}}(a+3c)
\end{eqnarray}

From Eqs (\ref{eq:SU3lim}) it follows that $SU(3)$-singlet resonances do
not
affect the final formulae since in the $SU(3)$ limit the $SU(3)$-singlet
state is not produced through an FSI-unmodified weak process.
Solution of Eqs (\ref{eq:FSIforb0}) for SU(3)-symmetric input amplitudes is
\begin{eqnarray}
\label{eq:FSIsol0}
 \tilde{B} & = & b \nonumber \\
 A-2B & = & a-2b \nonumber \\
 \tilde{A} + B & = & a+2b \nonumber \\
B+C-\tilde{C}-E-3F & = & b \nonumber \\
 \tilde{C}+F & = & c + T^{(0)}_2(c+\frac{a}{3})
\end{eqnarray}

A look at how resonance-induced FSI (Fig. 3 and its counterparts)
generate FSI-modified diagrams from the input $(a)$, $(b)$, $(c)$
amplitudes confirms that diagrams of type $(a)$ and $(b)$ cannot
actually be generated. Thus, we must have

\begin{eqnarray}
\label{eq:solution}
A=\tilde{A} &=& a \nonumber \\
B=\tilde{B} &=& b \nonumber \\
C-E-2F      &=& \tilde{C}+F
\end{eqnarray}

From Eqs (\ref{eq:FSIsol1},\ref{eq:FSIsol0},\ref{eq:solution})  we see
that in the case when
strong interactions exhibit $SU(3)$ symmetry, i.e. when the
intermediate $I=1$ and $I=0$ octet scalar resonances
are degenerate and couple to $PP$ with the same strength
$\kappa ^{I=0}_S = \kappa ^{I=1}_S = \kappa _S$
so that
\begin{eqnarray}
\label{eq:deg1}
\delta ^{I=0}_2 &=& \delta ^{I=1}_1 \equiv \delta \nonumber \\
T^{(0)}_2 &=& T^{(1)}_1
\end{eqnarray}
one obtains
\begin{eqnarray}
\label{eq:deg2}
E+F&=& \frac{1}{2}(T^{(1)}_1-T^{(0)}_2)(c+\frac{a}{3}) = 0 \nonumber \\
C+E=\tilde{C}+F&=& c+(\cos \delta \exp i\delta -1)(c+\frac{a}{3})
\end{eqnarray}

Thus, physically measurable amplitudes {\em may be described} with
vanishing
effective penguin amplitudes $E$, $F$ and $SU(3)$-symmetric exchange
amplitudes $\tilde{C} = C$.

In reality, in the energy region below $1500$ $MeV$
the scalar resonance sector
exhibits peculiar $SU(3)$-symmetry  breaking
\cite{Torn,Torn95}.
The masses and couplings of physical resonances are known to be
different
from those expected when these resonances are assigned a simple
$q\overline{q}$ structure.
Descriptions of these resonances as $qq\overline{q}\overline{q}$
states have been proposed. Despite years of intensive efforts the
questions related to the nature of these states have not been
settled as yet.
The strangeness $S=0$ sector under consideration in this section is
particularly troublesome
as exhibited by conflicting interpretations of the
$a_0(980)$, $f_0(980)$, and $f_0(1300)$ states \cite{other}.
One should therefore expect that
in the $D$-mass region around $1870$ $MeV$ the situation is also
complicated
(see also \cite{Don80}).
A detailed analysis of the nature of resonances affecting
two-meson interactions at this energy is clearly
far beyond the scope of this paper. For our purposes the crucial point
is the relative size of amplitudes corresponding
to diagrams of Fig. 2 and Fig. 4.

\begin{picture}(260,100)

\put(20,20){\begin{picture}(70,55)
\put(0,55){\line(2,-1){30}}
\put(30,40){\vector(1,0){5}}
\put(35,40){\line(1,0){5}}
\put(40,40){\line(2,1){30}}

\put(0,10){\line(2,1){30}}
\put(30,25){\line(1,0){5}}
\put(40,25){\vector(-1,0){5}}
\put(70,10){\line(-2,1){30}}

\put(0,45){\line(2,-1){30}}
\put(30,30){\line(1,0){5}}
\put(40,30){\vector(-1,0){5}}
\put(40,30){\line(2,1){30}}

\put(0,0){\line(2,1){30}}
\put(30,15){\vector(1,0){5}}
\put(40,15){\line(-1,0){5}}
\put(40,15){\line(2,-1){30}}

\end{picture}}

\put(140,20){\begin{picture}(70,55)
\put(0,55){\line(2,-1){30}}
\put(30,40){\vector(1,0){5}}
\put(35,40){\line(1,0){5}}
\put(40,40){\line(2,1){30}}

\put(0,10){\line(2,1){30}}
\put(40,30){\vector(-2,-1){10}}
\put(40,30){\line(2,1){30}}

\put(0,45){\line(2,-1){30}}
\put(30,30){\line(2,-1){10}}
\put(70,10){\vector(-2,1){30}}

\put(0,0){\line(2,1){30}}
\put(30,15){\vector(1,0){5}}
\put(40,15){\line(-1,0){5}}
\put(40,15){\line(2,-1){30}}

\end{picture}}

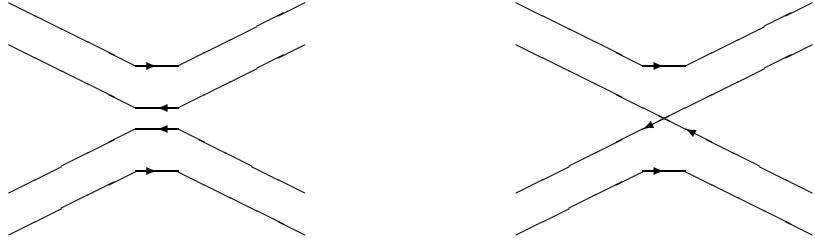
\put(20,0){Fig. 4 Contributions of four-quark intermediate states
to FSI}

\end{picture}

If diagrams of Fig. 4 do not contribute significantly to the mechanism
of physical resonance formation, as is the case in the unitarised
quark model (UQM) of T\"ornqvist \cite{Torn,Torn95}, one should expect that
the values of amplitudes $A,B$ may be taken from the $S=-1$ sector of $D^0$
decays or from
$D^+$ decays, and subsequently used in the $S=0$ sector.
On the other hand, the remaining parameters
of the quark-line approach needed
for the description of the $D^0 \rightarrow \pi \pi, K
\overline{K}$ etc. decays
may exhibit significant
$SU(3)$-breaking within and
between the I=0 and I=1
subsectors of the S=0 sector (and, of course, between the S=0 and S=-1
sectors) \cite{Torn}.
In the analysis of T\"ornqvist \cite{Torn95} the scalar meson sector
exhibits energy-dependent mixing of the two $I=0$ resonances. The value of
the mixing angle undergoes a fairly rapid change in the vicinity of the
$K\overline{K}$ threshold. Below $900$ $MeV$ the mixing is nearly ideal
while above $1.1$ $GeV$ one has nearly pure $SU(3)$ eigenstates.
The $f_0(1300)$ appears then as a near-octet resonance.
Although the analysis
of \cite{Torn95} stops at $1.6$ $GeV$ there are no reasons to expect a
qualitative change in the mixing angle when energy changes from $1.6$ to
$1.87$
$GeV$: The majority of $PP$ thresholds lie well below $1.6$ $GeV$.
Despite significant $SU(3)$ breaking incorporated into the UQM (i.e.
through realistic positions of thresholds) this model confirms essentially
that our use of pure $SU(3)$ eigenstates at $D^0$ energy is
justified.
$SU(3)$-breaking enters the $K_0$ matrix through the difference in
the real parts of the vacuum polarization functions $\Pi (s)$
\cite{Torn95}
which are different for each of the $I=1$ and two $I=0$ states.
Consequently, the $I=0$ and $I=1$ octet channels are affected
differently corresponding to strong $SU(3)$ breaking between $a_0(980)$
and $f_0(1300)$. In our simplified treatment this means that the
sizes of {\em effective} couplings in the $I=1$ and $I=0$ octet
channels are different.
As a result we should treat
the isospin amplitudes in the I=0 and I=1 sectors
as {\em independent} free parameters.

In terms of amplitudes with definite isospin the amplitudes of the
four measured $D^0 \rightarrow K\overline{K}, \pi \pi$ decays are given by
\begin{eqnarray}
\label{eq:isospin}
<K^+K^-|W|D^0> &=& \frac{1}{\sqrt{2}}({\cal K}_1+{\cal K}_0) \nonumber \\
<K^0\overline{K}^0|W|D^0> &=&
\frac{1}{\sqrt{2}}({\cal K}_1-{\cal K}_0) \nonumber \\
<\pi ^+\pi ^-|W|D^0> &=& \frac{1}{\sqrt{3}}({\cal P}_2+\sqrt{2}{\cal P}_0)
\nonumber \\
<\pi ^0\pi ^0|W|D^0> &=& \frac{1}{\sqrt{3}}(\sqrt{2}{\cal P}_2-{\cal P}_0)
\end{eqnarray}
where the amplitudes of definite isospin (specified by subscript) can be
expressed in terms of quark-line amplitudes as follows
\begin{eqnarray}
\label{eq:isospin2}
{\cal K}_1 &=& -\frac{1}{\sqrt{2}}(\tilde{A}+C+E) \nonumber \\
{\cal K}_0 &=& -\frac{1}{\sqrt{2}}(\tilde{A}+\tilde{C}+F) \nonumber \\
{\cal P}_2 &=& -\frac{1}{\sqrt{3}}(A+B) \nonumber \\
{\cal P}_0 &=& \sqrt{\frac{2}{3}}(-A+\frac{B}{2}-\frac{3}{2}(\tilde{C}+F))
= \frac{1}{\sqrt{6}}(A+B)+\sqrt{3}{\cal K}_0
\end{eqnarray}
Note that (apart from the contribution of the FSI-unmodified
$(a)$- and $(b)$-type amplitudes) the isospin $I=0$ $(1)$ amplitude
may be put in a one-to-one correspondence with the $\tilde{C}+F$
($C+E$) combination of quark-diagram amplitudes respectively.
Defining $Y=\frac{1}{\sqrt{2}}({\cal K}_1-{\cal K}_0)$ and
$X=\sqrt{2} {\cal K}_0$ the amplitudes of the four
considered decays acquire simple form given in Table 1,
where the corresponding experimental branching
ratios taken from \cite{PDG} are also displayed.
\\

Table 1. Theoretical amplitudes and branching ratios of the four
$D^0 \rightarrow KK,~\pi \pi$ decays measured.

\begin{center}
\begin{tabular}{l c c}
\hline
decay & amplitude & branching ratio in \% \\
\hline
$K^+~K^-$ & $X+Y$ & $0.454 \pm 0.029$ \\
$K^0~\overline{K}^0$ & $Y$ & $0.11 \pm 0.04$ \\
$\pi^+ \pi^-$ & $X$ & $0.159 \pm 0.012$ \\
$\pi^0 \pi^0$ & $-\frac{1}{\sqrt{2}}(A+B+X)$ & $0.088 \pm 0.023$ \\
\hline
\end{tabular}
\end{center}

The data of
Table 1 permit us to establish that
\begin{eqnarray}
\label{eq:modulae}
|X+Y|   &=& (4.51 \pm 0.14)*10^{-6}~GeV \nonumber \\
|Y|     &=& (2.21 \pm 0.40)*10^{-6}~GeV \nonumber \\
|X|     &=& (2.47 \pm 0.09)*10^{-6}~GeV \nonumber \\
|A+B+X| &=& (2.60 \pm 0.3) *10^{-6}~GeV
\end{eqnarray}
From the branching ratio of the $D^+ \rightarrow \pi ^+ \overline{K}^0$
decay (equal to $(2.74 \pm 0.29)*10^{-2}$ \cite{PDG}) described by
amplitude
$A+B$ one infers that
\begin{eqnarray}
\label{eq:D+}
|A+B|   &=& (1.35 \pm 0.07)*10^{-6}~GeV
\end{eqnarray}
Equations (\ref{eq:modulae}) and (\ref{eq:D+}) show that, contrary to the
conclusions of ref.\cite{ChCh92}, it is still possible to keep
$SU(3)$-symmetry in factorization
amplitudes $A,B$ provided one {\em correctly} describes final
state interactions
( $SU(3)$-symmetry of factorization amplitudes was used when writing
the last equality in Eq.(\ref{eq:isospin2})).
In particular, even if the data were consistent with $Y=0$ (and
$|X+Y|=|X|\approx 2.5 * 10^{-6}~GeV$),
i.e. if the amplitudes were $SU(3)$-symmetric, we
would still have to conclude from the values of
$|A+B|$, $|X|$, and $|A+B+X|$ that the relative phase of $A+B$ and $X$
must be close to $90^o$.
Since in the quark-diagram approach $X = -A-\tilde{C}-F$, it follows
that the relative phase of $\tilde{C}+F$ and $A$, $B$ must be significant,
in agreement with the message of this paper.
Nonzero phase of $\tilde{C}+F$ is a direct result of inelasticity
in FSI.

The $SU(3)$-breaking $Y$ amplitude is expressed through
quark-diagram amplitudes as $Y = -\frac{1}{2}(C+E-\tilde{C}-F)$.
Large observed size of $Y$ means that, when interpreted in terms of
quark-diagram amplitudes, the data can be described
either by a strong breaking of SU(3)-symmetry in $W$-exchange
amplitude
($C \neq
\tilde{C}$ and
$E-F \approx 0$)
or by a large contribution from effective long-range penguins ($E-F \ne 0$),
or both.
One has to keep in mind, however, that in the $SU(3)$-symmetry breaking
case the combination $C+E$ (determined from the $I=1$ sector) cannot be
used in the $I=0$ sector (i.e. in Eq.(\ref{eq:solution})): The diagrammatic
approach suggests more symmetry between the $I=0$ and $I=1$ sectors than
is actually present.

\section{Conclusions}
We have studied in some detail how inelastic coupled-channel rescattering
effects (and, in particular, $q \overline{q}$ resonance formation in the
final state)
modify the input weak amplitudes of the quark-line diagrammatical approach.
Through an explicit calculation it has been demonstrated that such
coupled-channel effects lead to the appearance of nonzero
relative phases between various quark diagrams, thus invalidating
the way in which final-state interactions were incorporated
into the diagrammatical approach in the past.
The case of $SU(3)$-symmetry breaking in Cabibbo once-forbidden
$D^0$ decays has been also discussed.
It has been shown that data may be described when
inelastic final-state interactions (which must be $SU(3)$-breaking as well)
are
introduced. On the other hand, contrary to statements in literature, the
data do not {\em require}
$SU(3)$-symmetry breaking in factorization amplitudes.

\section{Acknowledgments}
This research has been supported in part by Polish Committee for
Scientific Research Grant No. 2 P03B 231 08.


\begin{thebibliography}{99}

\bibitem {Chau86} L.-L. Chau and H.-Y. Cheng, Phys.Rev.Lett. 56, 1655
(1986).
\bibitem {Chau87} L.-L. Chau and H.-Y. Cheng, Phys. Rev. D36, 137, (1987).
\bibitem {Chau89} L.-L. Chau and H.-Y. Cheng, Phys. Lett. B222, 285 (1989).
\bibitem {BWS} M.Bauer, B. Stech and M. Wirbel, Z. Phys. C34, 103 (1987).
\bibitem {Lee} D. Lee, Phys. Lett. B275, 469 (1992).
\bibitem {Lipkin} H.J. Lipkin, Phys. Rev. Lett. 44, 710 (1980);
Phys. Rev. Lett. 46, 1307 (1981).
\bibitem {SorD23} C. Sorensen, Phys. Rev. D23, 2618 (1981).
\bibitem {Kamal} A.N. Kamal and E.D. Cooper, Z. Phys. C8, 67 (1981).
\bibitem {Don86} J.F. Donoghue, Phys. Rev. D33, 1516 (1986).
\bibitem {Chau83} L.-L. Chau, Phys. Rep. 95, 1 (1983).
\bibitem {ChCh92} L.-L. Chau and H.-Y. Cheng, Phys. Lett. B280, 281 (1992).
\bibitem {Hinchcli} I. Hinchcliffe and Th.A. Kaeding, LBL-35892 preprint.
\bibitem {Sorensen} C. Sorensen, Phys. Rev. D24, 2976 (1981).
\bibitem {Watson} K.N. Watson, Phys. Rev. 95, 228 (1954).
\bibitem {Babelon} O. Babelon et al., Nucl. Phys. B113, 445 (1976).
\bibitem {Chau89a} L.-L. Chau and H.-Y. Cheng, Phys. Rev. D39, 2788 (1989).
\bibitem {Don80} J.F.Donoghue and B.R.Holstein, Phys. Rev. D21, 1334
(1980).
\bibitem {Torn} N.A.T\"ornqvist, Phys. Rev. Lett. 49, 624 (1982);
Acta Phys. Pol. B16, 503 (1985); in {\em The Hadron Mass Spectrum},
Proceedings of the Rheinfels Workshop, St. Goar, Germany, 1990,
edited by E. Klempt and K. Peters, Nucl. Phys. B (Proc. Suppl.) Vol. 21
(1991).
\bibitem {Torn95} N.A.T\"ornqvist, Z.Phys. C68, 647 (1995)
and papers cited therein.
\bibitem {other} D.Morgan and M.R.Pennington, Phys.Rev.D48, 1185(1993);
D48, 5422 (1993); N.N.Achasov and G.N.Shestakov, Phys.Rev. D49, 5779
(1994); R.Kami\'nski, L.Le\'sniak, and J.-P.Maillet, Phys. Rev. D50,
3145 (1994).
\bibitem {PDG} Particle Data Group, {\em Review of Particle Properties},
Phys. Rev. D50, 1173 (1994).

\end{thebibliography}
\end{document}